\documentclass{llncs}

\usepackage{graphicx}
\usepackage{wrapfig}
\usepackage{subcaption}
\usepackage{paralist}

\usepackage{courier}
\usepackage{listings}
\usepackage{etoolbox}
\makeatletter
\pretocmd\lst@makecaption{\noindent{\rule{\linewidth}{2pt}}}{}{}
\makeatother

\usepackage{algorithm}
\usepackage{algorithmicx}
\usepackage[noend]{algpseudocode}

\usepackage{pgf}
\usepackage{tikz}
\usepackage{pgfplots}
\usetikzlibrary{matrix,trees,arrows,fit,shapes,positioning}
\usetikzlibrary{decorations.pathmorphing,decorations.pathreplacing,calc} % required in the preamble

\newboolean{showcomments}
\setboolean{showcomments}{true} 
\ifthenelse{\boolean{showcomments}}
  {\renewcommand{\note}[2]{
	\fbox{\bfseries\sffamily\scriptsize#1}
    {\sf\small$\Rightarrow$\textit{#2}$\Leftarrow$}
   }
  }
  {\renewcommand{\note}[2]{}
  }

\usepackage{amsfonts}

\usepackage[T1]{fontenc}
\usepackage{makecell}
\usepackage{multirow}

\begin{document}

\title{Optimized Execution of Business Processes\\ on Blockchain}
\author{Luciano Garc\'ia-Ba\~nuelos\inst{1} \and Alexander Ponomarev\inst{2}\and \\
 Marlon Dumas\inst{1} \and Ingo Weber\inst{2,3}}
\institute{
University of Tartu, Estonia\\
\email{\{luciano.garcia, marlon.dumas\}@ut.ee}
\and
Data61, CSIRO, Sydney, Australia\\
\email{\{alex.ponomarev, ingo.weber\}@data61.csiro.au}
\and
School of Computer Science \& Engineering, UNSW, Sydney, Australia\\
}
\maketitle

\begin{abstract}
Blockchain technology enables the execution of collaborative business processes involving untrusted parties without requiring a central authority. Specifically, a process model comprising tasks performed by multiple parties can be coordinated via smart contracts operating on the blockchain. The consensus mechanism governing the blockchain thereby guarantees that the process model is followed by each party. However, the cost required for blockchain use is highly dependent on the volume of data recorded and the frequency of data updates by smart contracts. This paper proposes an optimized method for executing business processes on top of commodity blockchain technology. The paper presents a method for compiling a process model into a smart contract that encodes the preconditions for executing each task in the process using a space-optimized data structure. The method is empirically compared to a previously proposed baseline by replaying execution logs, including one from a real-life business process, and measuring resource consumption.
\end{abstract}

% !TEX root = ../paper.tex
\section{Introduction}

Blockchain is commonly known as the technology underpinning bitcoin, but its potential applications go well beyond enabling digital currencies. Blockchain enables
an evolving set of parties to maintain a safe, permanent, and tamper-proof ledger of transactions without a central authority~\cite{UKBlockchainReport}. A key feature of this technology is that transactions are not recorded centrally. Instead, each party maintains a local copy of the ledger. The ledger is a linked list of blocks, each comprising a set of transactions. Transactions are broadcasted and recorded by each participant in the blockchain network. When a new block is proposed, the participants in the blockchain network collectively agree upon a single valid copy of this block according to a consensus mechanism. Once a block is collectively accepted, it is practically impossible to change it or remove it. Hence, a blockchain can be conceived as a replicated append-only transactional data store, which can serve as a substitute for a centralized register of transactions maintained by a trusted authority. 
Modern blockchain platforms such as Ethereum\footnote{\url{https://www.ethereum.org/} -- last accessed 4/12/2016} additionally offer the possibility of executing user-defined scripts on top of a blockchain when certain transactions take place. These so-called \emph{smart contracts} allow parties to encode business rules on the blockchain in a way that inherits from its tamper-proof properties, meaning that the correct execution of smart contracts is guaranteed by the protocols that ensure the integrity of the blockchain.

Blockchain technology opens manifold opportunities to redesign collaborative business processes such as supply chain and logistics processes~\cite{Dumas2016blockchain}. Traditionally, such processes are executed by relying on trusted third-party providers such as Electronic Data Interchange (EDI) hubs or escrows. This centralized architecture creates entry barriers and hinders process innovation. Blockchain enables these processes to be executed in a peer-to-peer manner without delegating trust to central authorities nor requiring mutual trust between each pair of parties.

Previous work~\cite{WeberBPM2016} has demonstrated the feasibility of executing collaborative business processes on a blockchain platform by transforming a collaborative process model into a smart contract serving as a template. From this template, instance-specific smart contracts are then spawned to monitor or execute each instance of the process. 
This initial proof-of-concept architecture has put into evidence the need to optimize resource usage. Indeed, the cost of blockchain technology is highly sensitive to the volume of data recorded on the ledger and the frequency with which these data are updated by smart contracts. In order to make blockchain technology a viable alternative for executing collaborative business processes, it is necessary to minimize the size of the code, the data maintained in the smart contracts and the frequency of data writes.
%, while still providing verifiability.

This paper proposes an optimized method for executing business processes defined in the standard Business Process Model and Notation (BPMN) on top of commodity blockchain technology. Specifically, the paper presents a method for compiling a BPMN process model into a smart contract defined in the Solidity language -- a language supported by Ethereum and other major blockchain platforms. The idea of the method is to translate the BPMN process model into a minimized Petri net and to compile this Petri net into a Solidity smart contract that encodes the ``firing'' function of the Petri net using a space-optimized data structure. The scalability of this method is evaluated and compared to the method proposed in~\cite{WeberBPM2016} by replaying business process execution logs of varying sizes and measuring the amount of paid resources (called ``gas'' in the Ethereum jargon) spent to deploy and execute the smart contracts encoding the corresponding business process models. 
Besides artificial models and logs, our experiments utilize a real-world process execution log with over 5,000 traces.

The rest of the paper is organized as follows. Section~\ref{sec:background} introduces blockchain technology and discusses previous work on blockchain-based process execution. Section~\ref{sec:bpmn2petrinets} presents the translation of BPMN models to Petri nets and the compilation of the latter to Solidity code. Section~\ref{sec:evaluation} presents the experimental evaluation. Finally, Section~\ref{sec:conclusion} draws conclusions and outlines future work.
% !TEX root = ../paper.tex
\section{Background and Related Work}
\label{sec:background}

This section introduces blockchain technologies and its performance costs, as well as previous work on the use of blockchain for collaborative process execution.

\subsection{Blockchain Technology}

The term blockchain refers both to a network and a data structure. As a data structure, a blockchain is a linked list of blocks, each containing a set of transactions. 
%The list starts with a \emph{genesis} block. 
Each block is cryptographically chained to the previous one by including its hash value and a cryptographic signature, in such a way that it is impossible to alter an earlier block without re-creating the entire chain since that block.
The data structure is replicated across a network of machines. 
Each machine holding the entire replica is called a \emph{full node}.
In \emph{proof-of-work blockchains,} such as Bitcoin and Ethereum, some full nodes play the role of \emph{miners}: 
they listen for announcements of new transactions, broadcast them, and try to create new blocks that include previously announced transactions.
Block creation requires solving a computationally hard cryptographic puzzle.
%, such as varying some parameters in the block to find a combination that results in a hash value smaller than a target. 
%At the time of writing, the Bitcoin blockchain requires a new block's hash value to have at least $17$ leading zeros (although that is not a sufficient criterion, as the difficulty is more fine-grained than just then number of leading zeros).
Miners race to find a block that links to the previous one and solves the puzzle. The winner is rewarded with an amount of new crypto-coins and the transaction fees of all included transactions.
%(which is how coins are ``minted'') 

%Several blockchain networks are based on open-source software, so it is relatively easy to start one's own blockchain: deploy the software and start a new chain with a new genesis block.
%The operators of a new blockchain can decide whether to run the new network as a \emph{public} one, meaning that anyone can join as a full node (with full read access to the entire  history), a miner, or just a user submitting transactions. Alternatively, a \emph{private blockchain} is a network where only authenticated and authorized users can join in any capacity. 
%In a \emph{permissioned blockchain,} some users are authorized to read, others to transact, and yet others to mine new blocks.
%For our experiments, we create private blockchains to avoid transaction fees and other costs of using public networks.

The first generation of blockchains were limited to the above functionality with minor extensions. %, such as enriching transactions with data. 
%Such data can, e.g., provide a remittance advice with the actual transfer of assets itself, but can also represent arbitrary other information, such as an MP3 file or an image. Since the data is replicated thousand-fold around the world, storage fees are relatively high.
The second generation added the concept of \emph{smart contracts}: scripts that are executed whenever a certain type of transaction occurs and which may themselves read and write from the blockchain.
Smart contracts allow parties to enforce that whenever a certain transaction takes place, other transactions also take place. Consider for example a public registry for land titles. Such a registry can be implemented as a blockchain that records who owns which property at present. Selling a property can be implemented as a transaction, cryptographically signed by both the vendor and the buyer. By attaching a smart contract to sales transactions, it is possible to enforce that when a sale takes place, the corresponding funds are transfered, the corresponding tax is paid, and the land title is transferred in a single action. 
%It is also possible to enforce that only transactions with sufficient funds for tax and purchase get accepted.
This example illustrates how smart contracts can enforce the correct execution of collaborative processes.
%Thus, smart contracts enable the enforcement of (almost) arbitrary forms of interaction between mutually untrusting parties.

The \emph{Ethereum}~\cite{EthWhitePaper} blockchain treats smart contracts as first-class elements.
It supports a dedicated language for writing smart contracts, namely Solidity.
Solidity code is translated into bytecode to be executed on the so-called \emph{Ethereum Virtual Machine (EVM)}.
When a contract is deployed through a designated transaction, the cost depends on the size of the deployed bytecode~\cite{EthYellowPaper}.
A Solidity smart contract offers methods that can be called via transactions. In the above example, the land title registry could offer a method to read current ownership of a title, and another one for transferring a title.
When submitting a transaction that calls a smart contract method, the transaction has to be equipped with crypto-coins in the currency \emph{Ether}, in the form of \emph{gas}.
This is done by specifying a gas limit (e.g. 2M gas) and gas price (e.g., $10^{-8}$ Ether / gas), and thus the transaction may use up to gas limit $\times$ price (2M $\times 10^{-8}$ Ether = 0.02 Ether).
Ethereum's cost model is based on fixed gas consumption per operation~\cite{EthYellowPaper}, e.g., reading a variable costs 50 gas,  writing a variable 5-20K gas, and a comparison statement 3 gas. 
Data write operations are significantly more expensive than read ones. Hence, when optimizing Solidity code towards cost, it is crucial to minimize data write operations on variables stored on the blockchain. Meanwhile, the size of the bytecode needs to be kept low to minimize deployment costs.
%Market mechanisms of the network determine the accepted gas price, i.e., if the offered gas price is below the winning miner's limit, the transaction does not get included.
%If the transaction gets included, the winning miner receives the fee, i.e., offered gas price $\times$ actually consumed gas. 
%If the user-specified gas limit is insufficient to execute the smart contract method, the execution is terminated at an undefined point, usually with no effect except for paying the fee. Clients can test-execute the transaction on their local full node to determine a suitable gas limit.

%In our conceptual and implementation work, we made use of documentation from the Ethereum community such as the Ethereum White Paper~\cite{EthWhitePaper}.
%Note that none of these publications have undergone an academic peer review; they have, however, been available for community peer discussion.
%The Ethereum Yellow Paper~\cite{EthYellowPaper} describes the inner workings of Ethereum. It includes a detailed description of the EVM cost model on the bytecode level.
%An issue on Ethereum's github project was helpful in assessing the details of costing for 8-bit variables.

\subsection{Related Work}

In prior work~\cite{WeberBPM2016}, we proposed a method to translate a BPMN Choreography model into a Solidity smart contract, which serves as a factory to create choreography instances. From this factory contract, instance contracts are created by providing the participants' public keys. In the above example, an instance could be created to coordinate a property sale from a vendor to a buyer. Thereon, only they are authorized to execute restricted methods in the instance contract.
%In a BPMN choreography, activities represent message exchanges.
Upon creation, the initial activity(ies) in the choreography is/are enabled. 
When an authorized party calls the method corresponding to an enabled activity, the calling transaction is verified, and if successful, the method is executed and the state of the instance is updated, meaning that the executed activity is deactivated and subsequent activities are enabled. The set of enabled activities is determined by analyzing the gateways between the activity that has just been completed, and subsequent activities.

%\footnote{\url{https://ethereum.github.io/browser-solidity/\#version=soljson-v0.2.1+commit.91a6b35.js} -- we use the same version as in~\cite{WeberBPM2016} for comparability.}

The state of the process is captured by a set of Boolean variables, specifically one variable per task and one per incoming edge of each join gateway.
In Solidity, Boolean variables are stored as 8-bit unsigned integers, with $0$ meaning \texttt{false} and $255$ meaning \texttt{true}.\footnote{\url{https://github.com/ethereum/EIPs/issues/93} -- last accessed 29/11/2016\label{ftnote:stacking-vars}} Solidity words are 256 bits long. The Solidity compiler we use has an in-built optimization mechanism that concatenates up to 32 8-bit variables into a 256-bit word, and handles redirection and offsets appropriately.
Nevertheless, at most 8 bits in the 256-bit word are actually required to store the information -- the remaining are wasted.
This waste increases the cost of deployment and write operations. In this paper, we seek to minimize the variables required to capture the process state so as to reduce execution cost (gas).

In a vision paper~\cite{Hull:ICSOC:2016}, the authors argue that the data-aware business process modeling paradigm is well suited to model business collaborations over blockchains.
The paper advocates the use of the Business Artifact paradigm~\cite{Nigam:IBM-SJ:2003} as the basis for a domain-specific language for business collaborations over blockchains.
This vision however is not underpinned by an implementation and does not consider optimization issues.
%This vision is not underpinned by an implementation or experiments and hence not comparable to the optimization efforts here.
Similarly~\cite{Norta2015} advocates the use of blockchain to coordinate collaborative business processes based on choreography models, but without considering optimization issues.
%method to generate executable code from these models for execution on a blockchain platform.
%considering cost optimization issues, nor does it
Another related work~\cite{Frantz:ECAS:2016} proposes a mapping from a domain specific language for ``institutions'' to Solidity. This work also remains on a high level, and does not indicate a working implementation nor it discusses optimization issues.
A Master's thesis~\cite{Chalmers:MastersThesis:2015} proposes to compile smart contracts from the functional programming language Idris to EVM bytecode. According to the authors, the implementation has not been optimized. %, and some ``obvious'' optimization steps are discussed.

%One source from the BPM industry community deserves mentioning: a BPMLeader.com post by Rikken~\cite{Rikken:bpmleader.com:2015} is the earliest source that we know discussing the idea of considering smart contracts in the wider business process landscape -- albeit this discussion remains on a high level.

%\md{Ingo to write:}
%\begin{itemize}
%\item One paragraph to introduce blockchain platforms and how they operate (miners, proof of work). Talk briefly about open versus permissioned versus private blockchains - in the experiments we'll probably need this to explain that we ran on a private copy.
%\item Introduce Ethereum; introduce smart contracts (maybe given an example), introduce Solidity and explain the costs associated with keeping variables and writing into variables in solidity (also introduce the term ``gas'').
%\item Introduce your previous work on blockchain-based execution of BPs; how the encoding of the process state (the marking is done) in this previous work, how the firing function is implemented in your previous work, and why this implementation ends up being expensive.
%\item Introduce other work on blockchain-based process execution - which are mostly white papers or conceptual papers that do not demonstrate how a process is executed on blockchain. In particular cite Hull's recent keynote paper on this topic.
%\end{itemize}
% !TEX root = ../paper.tex
\section{Method}
\label{sec:bpmn2petrinets}

\begin{wrapfigure}{l}{.5\textwidth}
\vspace{-8mm}
\centering
\includegraphics[scale=0.55]{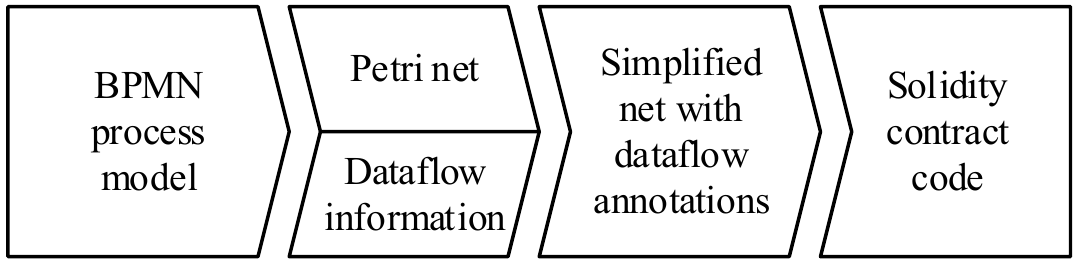}
\caption{\label{fig:chain}Chain of transformations}
\vspace{-8mm}
\end{wrapfigure}
Figure~\ref{fig:chain} shows the main steps of our method. The method takes as input a BPMN process model. The model is first translated into a Petri net. A dataflow analysis is applied to determine, where applicable, conditions that constrain the execution of each task. 
Next, reduction rules are applied to the Petri net to eliminate invisible transitions and spurious places. The minimized net is annotated with metadata extracted by the dataflow analysis. Finally, the minimized net is compiled into Solidity. Below, we discuss each step in detail.

\subsection{From BPMN to Petri nets}
\enlargethispage{\baselineskip}

\begin{figure}[b!]
\vspace*{-4mm}
\centering
\includegraphics[scale=0.65]{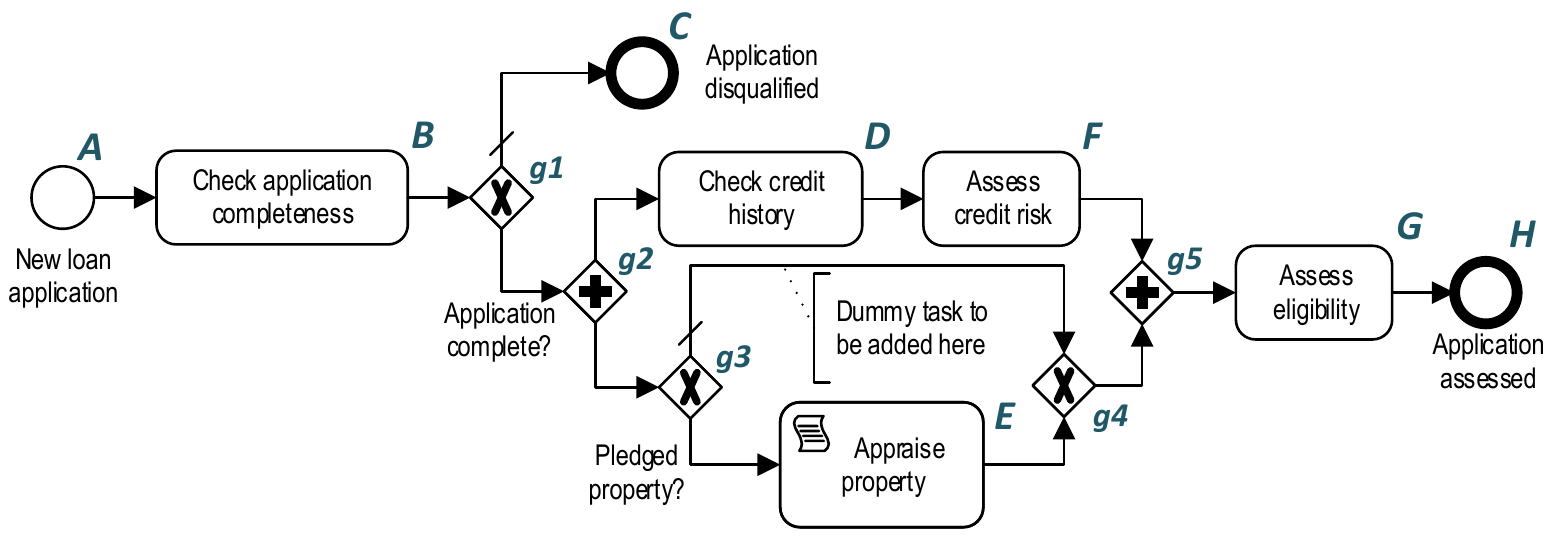}
\caption{\label{fig:loanproc:bpmn}Loan assessment process in BPMN notation}
\vspace*{-3mm}
\end{figure}

The proposed method takes as input a BPMN process model consisting of the following types of nodes: tasks, plain and message events (including start and end events), exclusive decision gateways (both event-based and data-based ones), merge gateways (XOR-joins), parallel gateways (AND-splits), and synchronization gateways (AND-joins). Figure~\ref{fig:loanproc:bpmn} shows a running example of BPMN model. Each node is annotated with a short label (e.g. $A, B, g1 \ldots$) for ease of reference.

\begin{figure}[t!]
\centering
\includegraphics[scale=0.65]{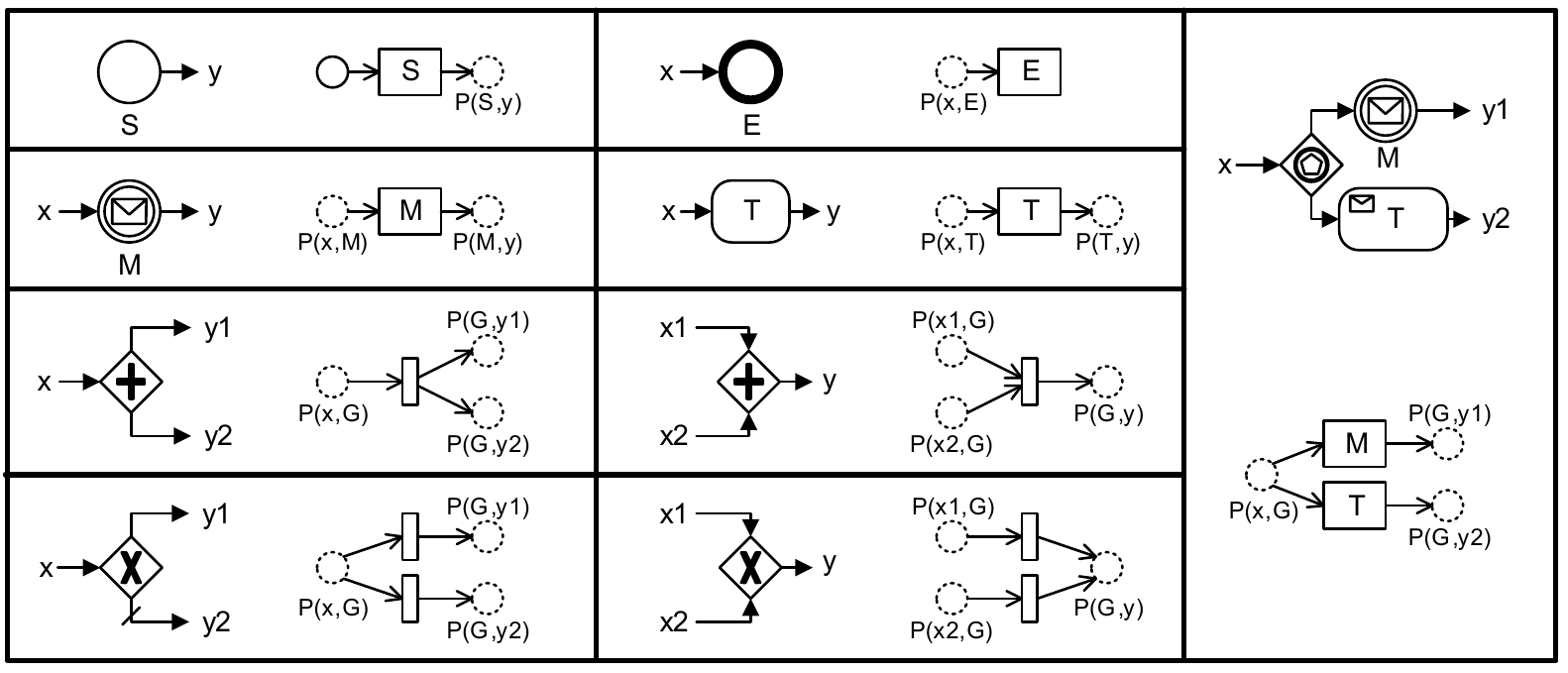}
\caption{\label{fig:bpmn2pnets:mapping}Mapping of BPMN elements  into Petri nets}
\vspace*{-3mm}
\end{figure}

\begin{figure}[b!]
\vspace*{-4mm}
\centering
\begin{tikzpicture}[thick,font={\fontsize{8pt}{12}\selectfont}]
\matrix[row sep=1.5mm, column sep=4mm] {
& & & & & \node (tau0p) {}; & \node (p3p) {}; & \node (cp) {}; \\
\node {\phantom{.}};\\
\node (p0p) {}; & \node (ap) {};  & \node (p1p) {}; & \node (bp) {};  & \node (p2p) {};& & & & \node (p5p) {}; & \node (dp) {};  & \node (p7p) {}; & \node (fp) {}; &  \node (p9p) {}; \\
& & & & & \node (tau1p) {}; & \node (p4p) {};  & \node (tau2p) {};  &&&&&& \node (tau5p) {}; & \node (p11p) {};  & \node (gp) {}; & \node (p12p) {}; &\node (hp) {}; \\
&&&&&&&&& \node (tau3p) {}; & \node (p00p) {}; & \node (tau00p) {};\\
&&&&&&& \node (p6p) {}; &&&&&& \node (p10p) {};\\
&&&&&&&& \node (tau4p) {};&\node (p8p) {};& \node (ep) {}; & \node (p01p) {}; & \node (tau01p) {};\\
};

\node[draw,circle] (p0) at (p0p) {};
\node[draw,fill,circle,inner sep=0,minimum width=1mm] at (p0p) {};
\node[draw,circle] (p1) at (p1p) {};
\node[draw,circle] (p2) at (p2p) {};
\node[draw,circle] (p3) at (p3p) {};
\node[draw,circle] (p4) at (p4p) {};
\node[draw,circle] (p5) at (p5p) {};
\node[draw,circle] (p6) at (p6p) {};
\node[draw,circle] (p7) at (p7p) {};
\node[draw,circle] (p8) at (p8p) {};
\node[draw,circle] (p9) at (p9p) {};
\node[draw,circle] (p10) at (p10p) {};
\node[draw,circle] (p11) at (p11p) {};
\node[draw,circle] (p12) at (p12p) {};
\node[draw,circle] (p00) at (p00p) {};
\node[draw,circle] (p01) at (p01p) {};

\node[draw] (a) at (ap) {A};
\node[draw] (b) at (bp) {B};
\node[draw] (c) at (cp) {C};
\node[draw] (d) at (dp) {D};
\node[draw] (e) at (ep) {E};
\node[draw] (f) at (fp) {F};
\node[draw] (g) at (gp) {G};
\node[draw] (h) at (hp) {H};

\node[draw, inner sep=0pt, minimum height=10pt, minimum width=2pt] (tau0) at (tau0p) {}; 
\node[draw, inner sep=0pt, minimum height=10pt, minimum width=2pt] (tau1) at (tau1p) {}; 
\node[draw, inner sep=0pt, minimum height=10pt, minimum width=2pt] (tau2) at (tau2p) {}; 
\node[draw, inner sep=0pt, minimum height=10pt, minimum width=2pt] (tau3) at (tau3p) {}; 
\node[draw, inner sep=0pt, minimum height=10pt, minimum width=2pt] (tau4) at (tau4p) {}; 
\node[draw, inner sep=0pt, minimum height=10pt, minimum width=2pt] (tau5) at (tau5p) {}; 
\node[draw, inner sep=0pt, minimum height=10pt, minimum width=2pt] (tau00) at (tau00p) {}; 
\node[draw, inner sep=0pt, minimum height=10pt, minimum width=2pt] (tau01) at (tau01p) {}; 

\draw[->] (p0) edge (a) (a) edge (p1) (p1) edge (b) (b) edge (p2) (p2) edge (tau0) (p2) edge (tau1)
	(tau0) edge (p3) (tau1) edge (p4) (p3) edge (c) (p4) edge (tau2) (tau2) edge (p5) (tau2) edge (p6)
	(p5) edge (d) (d) edge (p7) (p7) edge (f) (f) edge (p9)
	(p6) edge (tau3) (p6) edge (tau4) (tau4) edge (p8) (p8) edge (e) 
	(tau3) edge (p00) (p00) edge (tau00) (tau00) edge (p10) (e) edge (p01) (p01) edge (tau01) (tau01) edge (p10)
	(p9) edge (tau5) (p10) edge (tau5) (tau5) edge (p11) (p11) edge (g) (g) edge (p12) (p12) edge (h)
;

\draw[dashed,color=red] ($(tau0.north west)+(-3pt,3pt)$) rectangle ($(c.south east)+(3pt,-3pt)$);
\draw[dashed,color=red] ($(tau1.north west)+(-3pt,3pt)$) rectangle ($(tau2.south east)+(3pt,-3pt)$);
\draw[dashed,color=red] ($(tau3.north west)+(-3pt,3pt)$) rectangle ($(tau00.south east)+(3pt,-3pt)$);
\draw[dashed,color=red] ($(tau4.north west)+(-3pt,3pt)$) rectangle ($(e.south east)+(3pt,-3pt)$);
\draw[dashed,color=red] ($(tau5.north west)+(-3pt,3pt)$) rectangle ($(g.south east)+(3pt,-3pt)$);
\draw[dashed,color=red] ($(e.north west)+(-3pt,3pt)$) rectangle ($(tau01.south east)+(3pt,-3pt)$);

\end{tikzpicture}
\caption{\label{fig:loanproc:fullnet}Petri net derived from example BPMN model}
\vspace*{-3mm}
\end{figure}
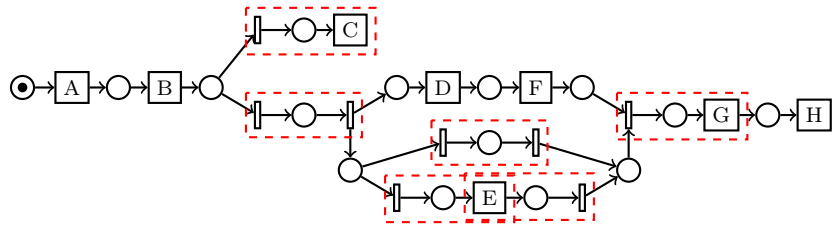

The BPMN process model is first translated into a Petri net using the transformation rules defined in~\cite{DijkmanDO08}, which are presented in Figure~\ref{fig:bpmn2pnets:mapping}.
Figure~\ref{fig:loanproc:fullnet} depicts the Petri net derived from the running example. The tasks and events in the BPMN model are encoded as labeled transitions (A, B, ...). Additional transitions without labels (herein called $\tau$ transitions) are introduced by the transformation to encode gateways as per the rules in Figure~\ref{fig:bpmn2pnets:mapping}.

%Also, since the workflow nets produced by this transformation have a property known as free-choiceness, if the generated workflow net fulfills the above two properties, it can be shown that there can be at most one token in a given place.

%This transformation allows us to check that the input model is sound, meaning that the resulting net has no deadlocks and is safe (one-bounded). 

%, by \A transition represents a system action. Visible transitions, i.e. transitions with a label, represent
%tasks or events in the model. 
%

%Each transition has a set of input places and a set
%of output places. At a given point in the execution of a Petri net, a place can
%hold a number of tokens. The distribution of tokens across places on the net is
%called a \emph{marking}. A transition is enabled and can ``fire'' when each of
%its input places has at least one token. When a transition fires, one token is
%removed from each input place and one token is added into each output place.

%\begin{definition}
%A tuple $(P,T,F,\lambda)$ is a \emph{labeled Petri net}, where $P$ is a set of \emph{places}, $T$ is a set of \emph{transitions}, with $P \cap T =\emptyset$,  $F \subseteq (P \times T) \cup (T \times P)$ is a set of arcs, and $\lambda: P \cup T \to \mathcal{L} \cup \{ \tau \}$ a labeling function.
%A \emph{net marking} $M: P \to \mathbb{N}_0$ is a function that associates a place $p \in P$ with a natural number (viz., place tokens). A \emph{net system} $N = (P,T,F,M_0)$ is a Petri net $(P,T,F)$ together with an \emph{initial marking} $M_0$.
%\end{definition}

The Petri net generated by this transformation is so-called a \emph{workflow net}. A workflow net has one source place (start), one sink place (end), and every transition is on a path from the start to the end. Two well-accepted behavioral correctness properties of workflow nets are
%\begin{itemize} 
%\item 
(i) \emph{Soundness:} starting from the marking with one token in the start place and no other token elsewhere (the \emph{initial marking}), it is always possible to reach the marking with one token in the end place and no other token elsewhere; and
(ii) \emph{Safeness:} starting from the initial marking, it is not possible to reach a marking where a place hold more than one token.
%\end{itemize}
%
These properties can be checked using available tools~\cite{DijkmanDO08}. Herein we assume that the Petri net resulting from the input BPMN model fulfills these properties. The third condition will allow us to encode the current marking in the net by associating a boolean to each place (is there a token in this place or not?) and this enables us to encode the marking as a bit array.

%In Figure~\ref{fig:loanproc:fullnet}, the left most place holds a token, represented by a
%filled circle. This means that transition {\sf A} is enabled. The firing of {\sf A} will remove 
%the token in its input place and then put back into its out place. The latter will enable transition {\sf B},
%so on and so forth.
%The sequence of firings will proceed until eventually a token is put before transition {\sf C}
%or transition {\sf H}, both corresponding to end events in the original process model. Note
%that neither {\sf C} nor {\sf H} have output places. When either of those transitions fire, 
%the token in their input node will be removed and no token will be added elsewhere in the net.
%This situation is interpreted as the end of the execution of the underlying process instance.

\subsection{Petri net reduction}

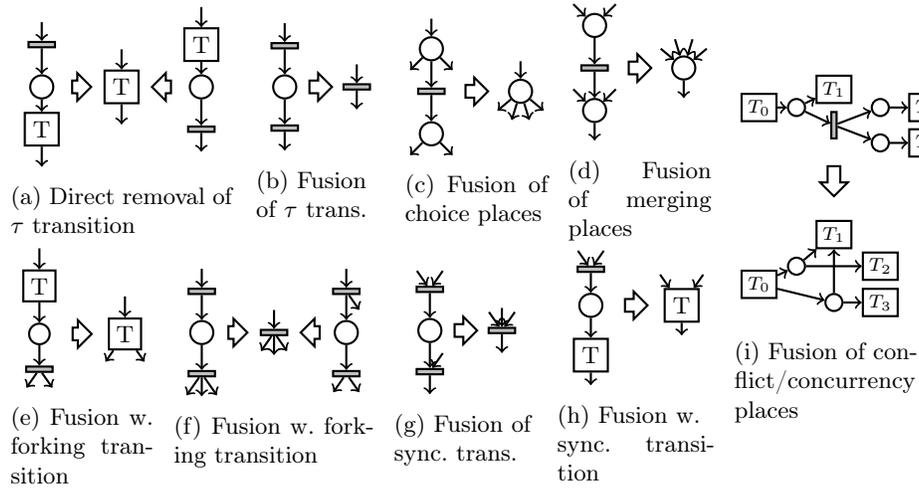
\begin{figure}[t!]
\begin{minipage}{.78\textwidth}
%% ---- 
\begin{minipage}{.31\textwidth}
\centering
\begin{tikzpicture}[thick]
\matrix[row sep=1mm, column sep=1.5mm] {
\node[draw,fill=lightgray, inner sep=0pt, minimum height=2pt, minimum width=10pt] (tau1) {};
&&&&\node[draw] (t3) {T};\\
\node[draw, circle] (p01) {}; &
\node[draw, inner sep=2pt,shape=single arrow, single arrow head extend=1mm, minimum height=2.5mm] {};&
\node[draw] (t2) {T}; &
\node[draw, inner sep=2pt,shape=single arrow, single arrow head extend=1mm, minimum height=2.5mm,rotate=180] {};&
\node[draw, circle] (p02) {};\\
\node[draw] (t1) {T};&&&&
\node[draw,fill=lightgray, inner sep=0pt, minimum height=2pt, minimum width=10pt] (tau2) {};\\
};
\draw[->] (tau1) edge (p01) (p01) edge (t1) (t1) edge ++(-90:5mm) (t2) edge ++(-90:5mm) (tau2) edge ++(-90:3mm)
	(t3) edge (p02) (p02) edge (tau2);
\draw[<-] (tau1) edge ++(90:3mm) (t2) edge ++(90:5mm) (t3) edge ++(90:5mm);
\end{tikzpicture}
\subcaption{Direct removal of $\tau$ transition}
\end{minipage}
\hspace{1mm}
\begin{minipage}{.18\textwidth}
\centering
\begin{tikzpicture}[thick]
\matrix[row sep=3mm, column sep=1.25mm] {
\node[draw,fill=lightgray, inner sep=0pt, minimum height=2pt, minimum width=10pt] (tau1) {};\\
\node[draw, circle] (p01) {}; &
\node[draw, inner sep=2pt,shape=single arrow, single arrow head extend=1mm, minimum height=3mm] {};&
\node[draw,fill=lightgray, inner sep=0pt, minimum height=2pt, minimum width=10pt] (tau3) {};\\
\node[draw,fill=lightgray, inner sep=0pt, minimum height=2pt, minimum width=10pt] (tau02) {};\\ 
};
\draw[->] (tau1) edge (p01) (p01) edge (tau02)
	(tau02) edge ++(-90:3mm) (tau3) edge ++(-90:3mm);
\draw[<-] (tau1) edge ++(90:3mm) (tau3) edge ++(90:3mm);
\end{tikzpicture}
\subcaption{Fusion\\of $\tau$ trans.}
\end{minipage}
\hspace{.5mm}
\begin{minipage}{.2\textwidth}
\centering
\begin{tikzpicture}[thick]
\matrix[row sep=2mm, column sep=2.5mm] {
\node[draw, circle] (p1) {};\\ 
\node[draw,fill=lightgray, inner sep=0pt, minimum height=2pt, minimum width=10pt] (tau) {}; &
\node[draw, inner sep=2pt,shape=single arrow, single arrow head extend=1mm, minimum height=3mm] {};&
\node[draw, circle] (p3) {};\\ 
\node[draw, circle] (p2) {};\\ 
};
\draw[->] (p1) edge (tau) (tau) edge (p2)
	(p1) edge ++(-135:4mm) (p1) edge ++(-45:4mm)
	(p2) edge ++(-45:4mm) (p2) edge ++(-135:4mm)
	;
\draw[<-] (p1) edge ++(90:4mm) (p3) edge ++(90:4mm);
	
\draw[->] (p3) edge ++(-40:4mm) (p3) edge ++(-73:4mm) (p3) edge ++(-107:4mm) (p3) edge ++(-140:4mm);
\end{tikzpicture}
\subcaption{Fusion of choice places}
\end{minipage}
\hspace{0.5mm}
\begin{minipage}{.2\textwidth}
\centering
\begin{tikzpicture}[thick]
\matrix[row sep=2mm, column sep=2.5mm] {
\node[draw, circle] (p1) {};\\ 
\node[draw,fill=lightgray, inner sep=0pt, minimum height=2pt, minimum width=10pt] (tau) {}; &
\node[draw, inner sep=2pt,shape=single arrow, single arrow head extend=1mm, minimum height=3mm] {};&
\node[draw, circle] (p3) {};\\ 
\node[draw, circle] (p2) {};\\ 
};
\draw[->] (p1) edge (tau) (tau) edge (p2);
\draw[<-]
	(p1) edge ++(135:4mm) (p1) edge ++(45:4mm)
	(p2) edge ++(45:4mm) (p2) edge ++(135:4mm)
	;
\draw[->] (p2) edge ++(-90:4mm) (p3) edge ++(-90:4mm);
	
\draw[<-] (p3) edge ++(40:4mm) (p3) edge ++(73:4mm) (p3) edge ++(107:4mm) (p3) edge ++(140:4mm);
\end{tikzpicture}
\subcaption{Fusion of merging places}
\end{minipage}

\begin{minipage}{.2\textwidth}
\centering
\begin{tikzpicture}[thick]
\matrix[row sep=2mm, column sep=2mm, inner sep=2.75pt] {
\node[draw] (t1) {T};\\
\node[draw, circle] (p01) {}; &
\node[draw, inner sep=2pt,shape=single arrow, single arrow head extend=1mm, minimum height=2.5mm] {};&
\node[draw] (t2) {T};\\
\node[draw,fill=lightgray, inner sep=0pt, minimum height=2pt, minimum width=10pt] (tau02) {};\\ 
};
\draw[->] (t1) edge (p01) (p01) edge (tau02);
\draw[<-] (t1) edge ++(90:5mm) (t2) edge ++(90:5mm);
\draw[->] (tau02) edge ++(-55:3mm) (tau02) edge ++(-125:3mm)
	(t2) edge ++(-55:4.5mm) (t2) edge ++(-125:4.5mm);
\end{tikzpicture}
\subcaption{Fusion w. forking transition}
\end{minipage}
\hspace{.5mm}
\begin{minipage}{.28\textwidth}
\centering
\begin{tikzpicture}[thick]
\matrix[row sep=3mm, column sep=1.5mm] {
\node[draw,fill=lightgray, inner sep=0pt, minimum height=2pt, minimum width=10pt] (tau1) {};&&&&
\node[draw,fill=lightgray, inner sep=0pt, minimum height=2pt, minimum width=10pt] (tau11) {};\\
\node[draw, circle] (p01) {}; &
\node[draw, inner sep=2pt,shape=single arrow, single arrow head extend=1mm, minimum height=2.5mm] {};&
\node[draw,fill=lightgray, inner sep=0pt, minimum height=2pt, minimum width=10pt] (tau3) {};&
\node[draw, inner sep=2pt,shape=single arrow, single arrow head extend=1mm, minimum height=2.5mm,rotate=180] {};&
\node[draw, circle] (p02) {}; \\
\node[draw,fill=lightgray, inner sep=0pt, minimum height=2pt, minimum width=10pt] (tau02) {};&&&&
\node[draw,fill=lightgray, inner sep=0pt, minimum height=2pt, minimum width=10pt] (tau12) {};\\ 
};
\draw[->] (tau1) edge (p01) (p01) edge (tau02) (tau11) edge (p02) (p02) edge (tau12);
\draw[<-] (tau1) edge ++(90:3mm) (tau3) edge ++(90:3mm) (tau11) edge ++(90:3mm);
\draw[->] (tau02) edge ++(-55:3.5mm) (tau02) edge ++(-125:3.5mm) (tau02) edge ++(-90:3mm)
	(tau3) edge ++(-55:3mm) (tau3) edge ++(-125:3mm) (tau3) edge ++(-90:3mm)
	(tau11) edge ++(-55:3mm) (tau12) edge ++(-55:3mm) (tau12) edge ++(-125:3mm);
\end{tikzpicture}
\subcaption{Fusion w. forking transition}
\end{minipage}
\hspace{.5mm}
\begin{minipage}{.2\textwidth}
\centering
\begin{tikzpicture}[thick]
\matrix[row sep=3mm, column sep=1.25mm] {
\node[draw,fill=lightgray, inner sep=0pt, minimum height=2pt, minimum width=10pt] (tau1) {};\\
\node[draw, circle] (p01) {}; &
\node[draw, inner sep=2pt,shape=single arrow, single arrow head extend=1mm, minimum height=3mm] {};&
\node[draw,fill=lightgray, inner sep=0pt, minimum height=2pt, minimum width=10pt] (tau3) {};\\
\node[draw,fill=lightgray, inner sep=0pt, minimum height=2pt, minimum width=10pt] (tau02) {};\\ 
};
\draw[->] (tau1) edge (p01) (p01) edge (tau02) (tau02) edge ++(-90:3mm) (tau3) edge ++(-90:3mm); 
\draw[<-] (tau1) edge ++(55:3mm) (tau1) edge ++(125:3mm) (tau02) edge ++(55:3mm)
	(tau3) edge ++(55:3mm) (tau3) edge ++(125:3mm) (tau3) edge ++(90:3mm);
%\draw[->] (tau02) edge ++(-55:3.5mm) (tau02) edge ++(-125:3.5mm) (tau02) edge ++(-90:3mm)
%	(tau3) edge ++(-55:3.5mm) (tau3) edge ++(-125:3.5mm) (tau3) edge ++(-90:3mm)
%	(tau11) edge ++(-55:3.5mm) (tau12) edge ++(-55:3.5mm) (tau12) edge ++(-125:3.5mm);
\end{tikzpicture}
\subcaption{Fusion of\\ sync. trans.}
\end{minipage}
\hspace{.5mm}
\begin{minipage}{.22\textwidth}
\centering
\begin{tikzpicture}[thick]
\matrix[row sep=2mm, column sep=2mm] {
\node[draw,fill=lightgray, inner sep=0pt, minimum height=2pt, minimum width=10pt] (tau) {};\\ 
\node[draw, circle] (p01) {}; &
\node[draw, inner sep=2pt,shape=single arrow, single arrow head extend=1mm, minimum height=3mm] {};&
\node[draw] (t2) {T};\\
\node[draw] (t1) {T};\\
};
\draw[->] (tau) edge (p01) (p01) edge (t1);
\draw[->] (t1) edge ++(-90:4mm) (t2) edge ++(-90:4mm);
\draw[<-] (tau) edge ++(55:3mm) (tau) edge ++(125:3mm)
	(t2) edge ++(55:5mm) (t2) edge ++(125:5mm);
\end{tikzpicture}
\subcaption{Fusion w.\\sync. transition}
\end{minipage}
\end{minipage}
%\hspace{0.5mm}
\begin{minipage}{.2\textwidth}
\centering
\begin{tikzpicture}[thick,font=\scriptsize]
\matrix[row sep=-1.5mm, column sep=1.25mm,ampersand replacement=\&, inner sep=0.8mm] {
\& \& \node[draw] (b) {$T_1$}; \\
\node[draw] (a) {$T_0$}; \& \node[draw, circle] (p1) {}; \& \& \node[draw, circle] (p2) {}; \& \node[draw] (c) {$T_2$};\\
\&\& \node[draw,fill=lightgray, inner sep=0pt, minimum height=10pt, minimum width=2pt] (tau) {}; \\
\&\&\& \node[draw, circle] (p3) {}; \& \node[draw] (d) {$T_3$};\\
\node[inner sep=0pt, minimum height=4mm] {}; \\
\& \& \node[draw, inner sep=2pt,shape=single arrow, single arrow head extend=1mm, minimum height=4mm,rotate=-90] {}; \\
\node[inner sep=0pt, minimum height=6mm] {}; \\
\& \& \node[draw] (bp) {$T_1$};\\
\node {\phantom{A}};\\
\& \node[draw, circle] (p01) {}; \& \& \node[draw] (cp) {$T_2$};\\
\node[draw] (ap) {$T_0$};\\
\& \& \node[draw, circle] (p02) {}; \& \node[draw] (dp) {$T_3$};\\
};
\draw[->] (a) edge (p1) (p1) edge (b) (p1) edge (tau) (tau) edge (p2) (tau) edge (p3) (p2) edge (c) (p3) edge (d);
\draw[->] (ap) edge (p01) (ap) edge (p02) (p01) edge (bp) (p02) edge (bp) (p01) edge (cp) (p02) edge (dp);
\end{tikzpicture}
\subcaption{Fusion of conflict/concurrency places}
\end{minipage}
\caption{\label{fig:reduction:rules}Toolkit of net reduction rules}
\end{figure}

The Petri net obtained from the previous step contains many $\tau$ transitions. If we consider each transition as an execution step, the number of steps required to execute this Petri net is unnecessarily high. 
%from a BPMN model captures the essence of the original model and
%is convenient for a number of analyses as outlined in~\cite{DijkmanDO08}. 
It is well-known that Petri nets with $\tau$ transitions can be reduced into smaller equivalent nets~\cite{Murata89}, under various notions of equivalence.
%if we adopt a well-accepted notion of behavioral equivalence known as 
%However, in most cases there is
%a smaller Petri net that captures the same behavior, if we consider a relaxed notion of equivalence
%As we discuss later, this notion of equivalence is acceptable in our setting. 
Here, we use the reduction rules presented in Figure~\ref{fig:reduction:rules}. 
Rules (a), (b), and (e)-(h) are fusions of series of transitions, 
whereas rules (c) and (d) are fusions of series of places. Rule(i) deals with $\tau$ transitions created by combinations of decision gateways and AND-splits.
These rules are designed so that the resulting net does not have any place that is both input and output of the same transition, as this would introduce infinite loops in the generated code. It can be proved that each of these reduction rules produces a Petri net that is \emph{weak trace equivalence} to the original one, i.e.\ it generates the same traces (modulo $\tau$ transitions) as the original one.

%We will consider that script tasks are used to inline Solidity snippets. All the other tasks are assumed to correspond with tasks that are externally called, e.g. from a DApp. Externally called tasks would typically be used to send some information to the in-blockchain process, that can later be used for further processing. In a nutshell, the translation of the BPMN model into Solidity code starts by (conceptually) generating a Petri net. The Solidity code, in turn, would simply implement the well known token game according to the behavior specified by the Petri net. Figure~\ref{fig:loanproc:net} presents a Petri net for the process model in Figure~\ref{fig:loanproc:bpmn}.

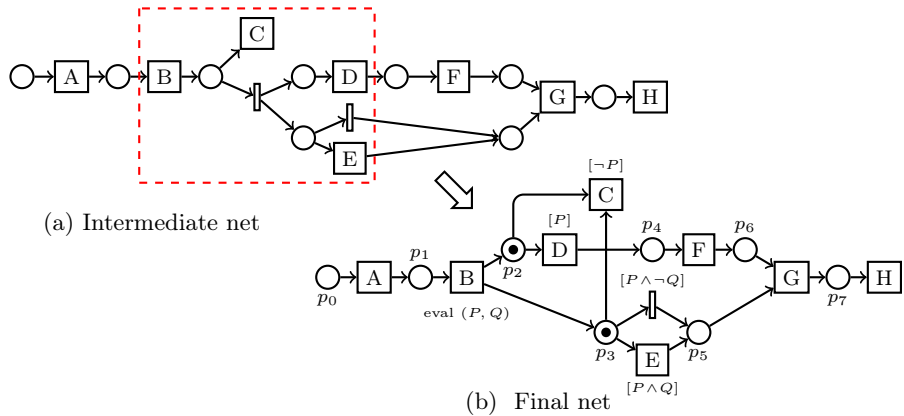
\begin{figure}[t!]
\begin{minipage}{.325\textwidth}
\vspace{-2.5cm}
\begin{tikzpicture}[thick,font={\fontsize{8pt}{12}\selectfont}]
\matrix[row sep=.55mm, column sep=4mm] {
&&&\node (xp) {}; && \node (cp) {}; \\
\node {\phantom{.}};\\
\node (p0p) {}; & \node (ap) {};  & \node (p1p) {}; & \node (bp) {};  & \node (p2p) {};& &  \node (p5p) {}; & \node (dp) {};  & \node (p7p) {}; & \node (fp) {}; &  \node (p9p) {}; \\
& & & & & \node (tau2p) {};  &&&&& & \node (gp) {}; & \node (p12p) {}; &\node (hp) {}; \\
&&&&&&& \node (tau3p) {};\\
&&&&&& \node (p6p) {}; &&&& \node (p10p) {};\\
&&&&&&&  \node (ep) {};\\
&&&&&&&&& \node[draw, inner sep=2pt,shape=single arrow, single arrow head extend=1mm, minimum height=6mm,rotate=-45] {}; \\
\\
};

\node[draw,circle] (p0) at (p0p) {};
\node[draw,circle] (p1) at (p1p) {};
\node[draw,circle] (p2) at (p2p) {};
\node[draw,circle] (p5) at (p5p) {};
\node[draw,circle] (p6) at (p6p) {};
\node[draw,circle] (p7) at (p7p) {};
\node[draw,circle] (p9) at (p9p) {};
\node[draw,circle] (p10) at (p10p) {};
\node[draw,circle] (p12) at (p12p) {};

\node[draw] (a) at (ap) {A};
\node[draw] (b) at (bp) {B};
\node[draw] (c) at (cp) {C};
\node[draw] (d) at (dp) {D};
\node[draw] (e) at (ep) {E};
\node[draw] (f) at (fp) {F};
\node[draw] (g) at (gp) {G};
\node[draw] (h) at (hp) {H};
\node (x) at (xp) {};

\node[draw, inner sep=0pt, minimum height=10pt, minimum width=2pt] (tau2) at (tau2p) {}; 
\node[draw, inner sep=0pt, minimum height=10pt, minimum width=2pt] (tau3) at (tau3p) {}; 

\draw[->] (p0) edge (a) (a) edge (p1) (p1) edge (b) (b) edge (p2)
	(p2) edge (c) (p2) edge (tau2)
	(tau2) edge (p5) (tau2) edge (p6)
	(p5) edge (d) (d) edge (p7) (p7) edge (f) (f) edge (p9)
	(p6) edge (tau3) 
	(p6) edge (e) (tau3) edge (p10) (e) edge (p10)
	(p9) edge (g) (p10) edge (g) (g) edge (p12) (p12) edge (h)
;

\draw[dashed,color=red] ($(x.north west)+(-6pt,6pt)$) rectangle ($(e.south east)+(3pt,-3pt)$);
\end{tikzpicture}
\vspace{-.75cm}
\subcaption{\label{fig:reduction:step1}Intermediate net}
\end{minipage}
%\caption{\label{fig:loanproc:onet:min}Incremental minimization of the Petri net}
\begin{minipage}{.5\textwidth}
\begin{tikzpicture}[thick,font={\fontsize{8pt}{12}\selectfont}]
\matrix[row sep=1.5mm, column sep=4mm] {
\node[minimum height=2cm] {};\\
&&&&\node (p2mp) {}; && \node (Cp){}; & & \\
\node {}; \\
&&&&\node (p2p) {}; & \node (Dp) {};  & & \node (p4p) {}; & \node (Fp) {};  & \node (p6p) {};\\
\node (p0p) {}; &\node (Ap) {}; &\node (p1p) {}; &\node (Bp) {}; & & & &&  & & \node (Gp) {}; & \node (p7p) {}; & \node (Hp) {};\\
&&&&&&& \node (taup) {}; \\
&&&&&&\node (p3p) {};&  & \node (p5p) {};\\
&&&&&&& \node (Ep) {}; \\
};
\node[draw] (A) at (Ap) {A};
\node[draw] (B) at (Bp) {B};
\node[draw] (C) at (Cp.center) {C};
%\node[draw,color=black!50!green] (D) at (Dp) {D};
%\node[draw,color=black!50!green] (E) at (Ep) {E};
\node[draw] (D) at (Dp) {D};
\node[draw] (E) at (Ep) {E};
\node[draw] (F) at (Fp) {F};
\node[draw] (G) at (Gp) {G};
\node[draw] (H) at (Hp) {H};
\node[draw, inner sep=0pt, minimum height=10pt, minimum width=2pt] (tau) at (taup) {};

\node[draw,shape=circle] (p0) at (p0p) {};
\node[draw,shape=circle] (p1) at (p1p) {};
\node[draw,shape=circle] (p2) at (p2p) {};
\node[draw,shape=circle] (p3) at (p3p) {};

\node[draw,shape=circle,fill,inner sep=0pt,text width=1mm] at (p2p) {};
\node[draw,shape=circle,fill,inner sep=0pt,text width=1mm] at (p3p) {};

\node[draw,shape=circle] (p4) at (p4p) {};
\node[draw,shape=circle] (p5) at (p5p) {};
\node[draw,shape=circle] (p6) at (p6p) {};
\node[draw,shape=circle] (p7) at (p7p) {};

\node[below=3pt] at (p0) {\scriptsize $p_0$};
\node[above=3pt] at (p1) {\scriptsize $p_1$};
\node[below=3pt] at (p2) {\scriptsize $p_2$};
\node[below=3pt] at (p3) {\scriptsize $p_3$};
\node[above=3pt] at (p4) {\scriptsize $p_4$};
\node[below=3pt] at (p5) {\scriptsize $p_5$};
\node[above=3pt] at (p6) {\scriptsize $p_6$};
\node[below=3pt] at (p7) {\scriptsize $p_7$};

\node[below=8pt] at (B) {\tiny eval $(P,Q)$};
\node[above=5pt] at (C) {\tiny $[\neg P]$};
\node[above=5pt] at (D) {\tiny $[P]$};
\node[above=3pt] at (tau) {\tiny $[P\!\land\!\neg Q]$};
\node[below=5pt] at (E) {\tiny $[P\!\land\!Q]$};

\draw[->]	(p0) edge (A) (A) edge (p1) (p1) edge (B)
		(B) edge (p2) (B) edge (p3)
		(p2) [rounded corners=1ex]-- (p2mp.center) --  (C) (p2) edge (D)
		(p3) edge (C) (p3) edge (E) (p3) edge (tau)
		(D) edge (p4)
		(E) edge (p5) (tau) edge (p5)
		(p4) edge (F) (F) edge (p6)
		(p5) edge (G) (p6) edge (G) (G) edge (p7) (p7) edge (H)
	;
\end{tikzpicture}
\vspace*{-8mm}
\subcaption{\label{fig:reduction:step2} Final net}
\end{minipage}
\caption{\label{fig:loanproc:net}Minimized Petri net for BPMN model in Fig.~\ref{fig:loanproc:bpmn}}
\vspace*{-4mm}
\end{figure}

The red dashes boxes in Figure~\ref{fig:loanproc:fullnet} show where the reduction rules can be applied. 
After applying the respective rules, we get the net shown in Figure~\ref{fig:reduction:step1}. At this point, we can still apply rule (i), which leads to the Petri net in Figure~\ref{fig:reduction:step2}.

\subsection{Dataflow analysis}

Some of the $\tau$ transitions generated by the BPMN-to-Petri net transformation correspond to conditions attached to decision gateways in the BPMN model. Since these $\tau$ transitions are removed by the reduction rules, we need to collect them back from the original model and re-attach them to transitions in the reduced net.
%, it becomes unclear where the evaluation of the conditions must be evaluated.
%Algorithm~\ref{alg:dataflow} to collect, from the original process model, the dataflow information that needs to be carried over to the minimized Petri net.
% inspired by the approach described in~\cite{DumasFMV05}, 
Algorithm~\ref{alg:dataflow} collects the conditions along each path between two consecutive tasks in a BPMN model, and puts them together into a conjunction. The algorithm performs a depth-first traversal starting from the start event. It uses two auxiliary functions: (i) \textproc{successorsOf}, which returns the set of direct successors of node $n$; and (ii) \textproc{cond}, which returns the condition attached to a sequence flow. Without loss of generality, we assume that every outgoing flow of a decision gateway has a condition attached to it (for a default flow, the condition is equal to the negation of the conjunction of conditions of its sibling flows). Also, we assume that any other sequence flow in the BPMN model is labeled with condition $true$ -- these $true$ labels can be inserted via pre-processing.

\begin{algorithm}
\begin{algorithmic}[1]
\State {\bf global} {\sf guards}: Map$\langle$Node $\mapsto$ Cond$\rangle = \emptyset$, {\sf visited}: Set$\langle$Node$\rangle$ = $\emptyset$
\Procedure{analyzeDataflow}{{\sf curr}: Node, {\sf predicate}: Cond}
	\State {\sf guards[curr] $\gets$ predicate}
	\State {\sf visited $\gets$ visited $\cup$ \{ current \}}
	\For{{\bf each} {\sf succ $\in$ \Call{successorsOf}{curr}} : succ $\not\in$ visited}
		\If {{\sf curr} is a Gateway}%\Comment{is curr a gateway?}
			\State{\sf \Call{analyzeDataflow}{succ, predicate $\land$ \textproc{cond}(curr, succ)}}
		\Else
			\State{\sf \Call{analyzeDataflow}{succ, $true$}}
		\EndIf
	\EndFor
\EndProcedure
\end{algorithmic}
\caption{\label{alg:dataflow}Dataflow analysis algorithm}% using label propagation
\end{algorithm}

\vspace*{-6mm}
Let us illustrate the algorithm assuming it traverses the nodes in the model of Figure~\ref{fig:loanproc:bpmn} in the following order:
$[A, B, g1, g2, g3, E, \dots]$.
%, where $g1-g3$ refer to gateways, enumerated from left to right. \iw{g1-g3 were undefined, please check my fix.}
In the first iteration, procedure \textproc{analyzeDataflow} sets {\sf guards = \{($A, true$)\}} in line 3 and proceeds until it calls itself recursively (line 9) with the only successor node of $A$, namely $B$. Note that {\sf predicate} is reset to $true$ in this recursive call. Something similar happens in the second iteration, where {\sf guard} is updated to $\{(A, true), (B, true)\}$. Again, the procedure is recursively called in line 9, now with node $g1$. This time {\sf guards} is updated to $\{(A, true), (B, true), (g1, true)\}$ but, since $g1$ is a gateway, the algorithm reaches line 7. There, the procedure is recursively called with {\sf succ} = $g2$ and  {\sf predicate} = $(true \land P)$, or simply $P$, where $P$ represents the condition 
``{\sf Application complete?}''. Since the traversal follows the sequence $[A, B, g1, g2, g3, E, \dots]$, it will eventually reach node $E$. When that happens, {\sf guards} will have the value 
 $\{(A, true), (B, true), (g1, true), (g2, P), (g3, P), (E, P \land Q)\}$, where $Q$ represents the condition
 ``{\sf Pledged property?}''. Intuitively, the algorithm would have propagated and combined the conditions $P$ and $Q$ while traversing the path between nodes $B$ to $E$. When the algorithm traverses $E$, the recursive call is done in line 9, where {\sf predicate} will be set to $true$, i.e. the predicate associated with $E$ will not be propagated further in the traversal. 
 %The depth-first search traversal finishes when all the nodes in the model have been visited.

The guards gathered by the above algorithm are then attached to the transitions in the minimized Petri net. In Figure~\ref{fig:reduction:step2} the guards collected by the algorithm are shown as labels next to the corresponding transitions. To avoid cluttering, $true$ guards are not shown. Note that the $\tau$ transition in the net in Figure~\ref{fig:reduction:step2} captures a situation where a task is skipped. Hence, this guard has to be included in the generated code. To this end, we insert a dummy (skip) task in the BPMN model to match each $\tau$ transition in the minimized net. The guard associated to a $\tau$ transition is then attached to its corresponding dummy task.
%To determine the guards associated with $\tau$ transitions, we will insert dummy tasks into the BPMN model \iw{In the BPMN model? Not in the minimized Petri net?} once we have determined where such transitions are needed, i.e. after the Petri net has been minimized.

For each transition in the minimized Petri net, we need to determine the set of conditions that need to be evaluated when it fires. To this end, for each transition we first compute the set of transitions that are reachable after traversing a single place, and then analyze the guards associated to such transitions. In our running example, we observe that transition $B$ can reach the set of transitions $\{C,D, E, \tau\}$ after traversing one place each. Hence, conditions $P$ and $Q$ need to be evaluated after task $B$ is executed, as hinted by the annotation ``eval $(P, Q)$'' below transition $B$ in Figure~\ref{fig:reduction:step2}. Thus, the evaluation of $P$ and $Q$ along with the net marking determine the set of transitions that will be enabled after $B$.

% !TEX root = ../paper.tex
\subsection{From minimized Petri net to Solidity}
\label{sec:petrinets2solidity}

In this step, we generate a Solidity smart contract that simulates the token game of the Petri net.  The smart contract uses two integer variables stored on the blockchain: one to encode the current {\sf marking} and the other to encode the value of the {\sf predicates} attached to transitions in the reduced net. Variable {\sf marking} is a bit array with one bit per place. This bit is set to zero when the place does not have a token, and one when the place holds a token. Note that thie requires that the placed in the net are deterministically ordered. This is done using their internal identifiers. To minimize space, the {\sf marking} is encoded as a 256-bits unsigned integer, which is the default word size in the EVM.

Consider the minimized Petri net in Figure~\ref{fig:reduction:step2}. 
Let us use the order indicated by the subscripts of the labels associated to the places of the net. The initial marking (i.e.\ the one with a token in $p_0$) is encoded as integer 1 (i.e. $2^0$). 
Hence, we initialize variable {\sf marking} with value 1 when an instance smart contract is created.
This marking enables transition $A$. The firing of $A$ removes the token from $p_0$ and puts a token in $p_1$. Token removal implemented via bitwise operations: {\sf marking = marking \& uint($\sim$1);}. Similarly, the addition of a token in $p_1$ (i.e. $2^1$ hence 2) is implemented via bitwise operations: {\sf marking = marking | 2;}.

Variable {\sf predicates} stores the current values of the conditions attached to the Petri net transitions. 
This variable is also an unsigned integer representing a bit array. As before, we first fix order the set of conditions in the process model, and associate one bit in the array per condition. For safety, particularly in the presence of looping behavior, the evaluation of {\sf predicates} is reset before storing the new value associated with the conditions that a given transition computes. For instance, transition $B$ first clears the bits associated with conditions $P$ and $Q$ (i.e. $2^0$ and $2^1$, respectively), and then stores the new values accordingly. 
%Again, we encode the {\sf predicates} bit array as an unsigned integer to save space.

When possible, an additional space optimization is achieved by merging variables {\sf marking} and {\sf predicates} into a single unsigned integer variable. The latter is possible if the number of places plus the number of predicates is at most 256.%, as EVM's default work size is 256 bits.%\footnote{Using smaller integers would be counterproductive as it would require a larger number of assembler operations, and hence more gas consumption, to unpack the value from and then pack it back to memory -- cf. link in Footnote~\ref{ftnote:stacking-vars}.}.
%EVM supports integers with sizes that are multiple of 8. However, 

%In principle, we need to generate a Solidity function for each transition of the Petri net. In order to apply some heuristic optimizations, we need to distiguish three cases. First, each transition associated with regular tasks gives rise to a public function. In this function, we first check if the transition is enabled or not: does the input places are marked? does the predicates evaluation corresponds with the transition guard? If the previous conditions hold we can proceed. The function then executes any Solidity code associated with the task (e.g. copying the values provided as parameters for the function call into contract values). Later, the function computes the set of predicates and updates the variable {\sf predicates} accordingly.

To illustrate how these variables are used to execute the process model, let us consider an excerpt of the Solidity smart contract associated with our running example (cf.\ Listing~\ref{listing:solidity}). The excerpt includes the code corresponding to transitions $B$, $E$ and the $\tau$ transition. Transition $B$ corresponds to task {\sf CheckApplication}. The corresponding function is shown in lines 5-18 in Listing~\ref{listing:solidity}. This task being a user task, this Solidity function will be called explicitly by an external actor, potentially with some data being passed as input parameters of the call (see line 5).
 %As it can be seen, in line 5 we hint the potential reception of input parameters. 
In line 6, the function checks if the marking is such that $p_2$ holds a token, i.e., if the current call is valid in that it \emph{conforms} to the current state of the process instance. 
If this is the case, the function will proceed
and execute the script task (cf. think of line 7 as a placeholder).
%, shows as placeholder line 7. 
Then the function evaluates predicates $P$ and $Q$ in lines 9-10. Note that the function does not immediately updates variable {\sf predicates} but stores the result in a local variable called {\sf tmpPred}, which we initialized in line 8.
In this way, we defer updating variable {\sf predicates} as much as possible (cf. line 42) to save gas ({\sf predicates} is a contract variable stored in the blockchain and writing to it costs 5000 gas). For the same reason, the new marking is computed in line 12 but the actual update to the respective contract variable {\sf marking} is deferred (cf. line 42).

\lstset{language=Java, basicstyle=\sffamily\scriptsize, literate={~} {$\sim$}{1},escapechar= @,
	emph={contract,function,returns,bool,uint,internal},emphstyle={\bfseries}
}
\lstset{numbers=left, numberstyle=\tiny, numbersep=5pt,float=tp}
\begin{lstlisting}[texcl,caption={\label{listing:solidity}Excerpt of Solidity contract}]
contract BPMNContract {
  uint marking = 1;
  uint predicates = 0;
  
  function CheckApplication( @\emph{-- input params --}@ ) returns (bool) {
    if (marking & 2 =@\,@= 2) { // is there a token in place $p_1$?
      // Task B's script goes here, e.g. copy value of input params to contract variables
      uint tmpPreds = 0;
      if ( @\emph{-- eval $P$ --}@ ) tmpPreds |= 1; // is loan application complete?
      if ( @\emph{-- eval $Q$ --}@ ) tmpPreds |= 2; // is the property pledged?
      step(
      	marking & uint(~2) | 12,       // New marking
	predicates & uint(~3) | tmpPreds // New evaluation for ``predicates''
      );
      return true;
    }
    return false;
  }
  
  function AppraiseProperty(uint tmpMarking) internal returns (uint) {
    // Task E's script goes here
    return tmpMarking & uint(~8) | 32;
  }
  
  function step(uint tmpMarking, uint tmpPredicates) internal {
    if (tmpMarking =@\,@= 0) { marking = 0; return; } // Reached a process end event!
    bool done = false;
    while (!done) {
    	// does $p_3$ have a token and does $P \land Q$ hold?
	if (tmpMarking & 8 =@\,@= 8 && tmpPredicates & 3 =@\,@= 3) {
	  tmpMarking = AppraiseProperty(tmpMarking);
	  continue;
	}
	// does $p_3$ have a token and does $P \land \neg Q$ hold?
	if (tmpMarking & 8 =@\,@= 8 && tmpPredicates & 3 =@\,@= 2) {
	  tmpMarking = tmpMarking & uint(~8) | 32;
	  continue;
	}
	...
	done = true;
    }
    marking = tmpMarking; predicates = tmpPredicates;
  }
  ... }
\end{lstlisting}

After executing $B$, if condition $P$ holds the execution proceeds with the possibility of executing $E$ or the $\tau$ transition. $E$ is a script task and can be executed immediately after $B$, if condition $Q$ holds, without any further interaction with external actors. For this reason, the Solidity function associated with task $E$ is declared as {\sf internal}. 
In the Solidity contracts that we create, all internal functions are tested for enablement, and if positive, executed. 
Specifically, the last instructions in any public function of the smart contract call a generic {\sf step} function (cf.\ lines 25-42 in Listing~\ref{listing:solidity}).
This function iterates over the set of internal functions, and executes the first activated one it finds, if any.
For instance, after executing $B$ there are tokens in $p_2$ and $p_3$. If $P \land Q$ holds, then the step function reaches line 31, where it calls function {\sf AppraiseProperty} corresponding to transition $E$. 
This function executes the task's script in line 21 and updates {\sf marking} in 22. After this, the control returns to line 32 in the {\sf step} function, which restarts the while loop. Once all the enabled internal functions are executed, we exit the while loop. In line 42, the {\sf step} function finally updates the contract variables.

%Lines 35-37 illustrate the handling of $\tau$ transitions. These transitions only move forward tokens, so there is no need for an explicit function. In this example, the $\tau$ transition moves the token from $p_3$ to $p_5$ (i.e. $2^3$ and $2^5$ respectively).

Algorithm~\ref{alg:template} sketches the functions generated for each transition in the minimized Petri net. Item 1 sketches the code for transitions associated to user tasks, while Item 2 does so for transitions associated to script tasks and $\tau$ transitions with predicates. For $\tau$ transitions without predicates, no function is generated, as these transitions only relay tokens (and this is done by the {\sf step} function).% in lines 25 to 43 of Listing 1.% based on the variable {\sf marking}.

% (which may be merged into one as per the optimization outlined earlier)
In summary, the code generated from the Petri net consists of a contract with the two variables {\sf marking} and {\sf predicates}, the functions generated as per Algorithm~\ref{alg:template} and the {\sf step} function. This smart contract offers one public function per user task (i.e. per task that requires external activation). This function calls the internal {\sf step} function, which fires all enabled transitions until it gets to a point where a new set of user tasks are enabled (or the instance has completed).
\vspace*{-8mm}
\enlargethispage{0.3\baselineskip}

\begin{algorithm}
\caption{\label{alg:template}Sketch of code generated  for each transition in the minimized net}
\scriptsize
\begin{compactenum}
\item For each transition associated to a user task, generate a public function with the following code:
\begin{itemize}
\item If task is enabled (i.e. check {\sf marking} and {\sf predicates}), then
\begin{enumerate}
\item {Execute the Solidity code associated with the task}
\item {If applicable, compute all predicates associated with this task and store the results in a local bit set, {\sf tmpPreds}}
\item Call {step} function with new marking and {\sf tmpPreds}, to execute all the internal functions that could become enabled
\item Return {TRUE} to indicate the successful execution of the task
\end{enumerate}
\item Return {FALSE} to indicate that the task is not enabled
\end{itemize} 
\item For each transition associated with a script task or $\tau$ transition that updates {\sf predicates}, generate an internal function with the following code:
\begin{enumerate}
\item {Execute the Solidity code associated with the task}
\item {If applicable, compute all predicates associated with this task and store the results in a local bit set, {\sf tmpPreds}}
\item Return the new marking and {\sf tmpPreds} (back to the {\sf step} function)
\end{enumerate}
%\item{For each $\tau$ transition that does not update {\sf predicates}, the {\sf marking} can be updated within {\sf step}}
\end{compactenum}
\end{algorithm}

% !TEX root = ../paper.tex
\section{Evaluation}
\label{sec:evaluation}

The goal of the proposed method is to lower the cost, measured in \emph{gas}, for executing collaborative business processes when executed as smart contracts on the Ethereum blockchain. Thus, we evaluate the output process contracts of our new translator comparatively against the previous translator's outputs.
The second question we investigate is that of throughput: is the approach sufficiently scalable to handle real workloads.
For sanity checking, we also check if the generated contracts can correctly discriminate conforming from non-conforming traces.
In this section, we start by introducing the datasets we use to these ends, followed by the experiment setup and methodology, and finally the experimental results.

\subsection{Datasets}

\begin{wraptable}{r}{.5\textwidth}
\vspace{-12mm}
\center\tiny
\begin{tabular}{|p{4.2em}|c|c|l|r|}
\hline
\textbf{Process} & \textbf{Tasks} & \textbf{GWs} & \textbf{Trace type} & \textbf{Traces}\\
\hline
Invoicing & 40 & 18 & Conforming & 5,316 \\
\hline
\multirow{2}{4.2em}{Supply chain} & \multirow{2}{*}{10} & \multirow{2}{*}{2}
& Conforming & 5 \\
\cline{4-5}
&&& Not conforming & 57 \\
\hline
\multirow{2}{4.2em}{Incident mgmt.} & \multirow{2}{*}{9} & \multirow{2}{*}{6}
& Conforming & 4 \\
\cline{4-5}
&&& Not conforming & 120 \\
\hline
\multirow{2}{4.2em}{Insurance claim} & \multirow{2}{*}{13} & \multirow{2}{*}{8}
& Conforming & 17 \\
\cline{4-5}
&&& Not conforming & 262 \\
\hline
\end{tabular}
\caption{Datasets used in the evaluation}
\label{tbl:datasets}
\vspace*{-6mm}
\end{wraptable}

For the evaluation purposes stated above, we draw on four datasets (i.e., logs and process models), statistics of which are given in Table~\ref{tbl:datasets}.
Three datasets are taken from our earlier work~\cite{WeberBPM2016}, the \emph{supply chain}, \emph{incident management}, and \emph{insurance claim} processes, for which we obtained process models from the literature and generated the set of conforming traces. Through random manipulation, we generated sets of non-conforming traces from the conforming ones.

%\iw{question for Marlon / Luciano: is this a real-world log? I thought so, but then you mentioned process mining platform and I'm not so sure anymore.}
The fourth dataset is stemming from a real-world invoicing process, which we received in the form of an event log with 65,905 events. This log was provided to us by the Minit process mining platform\footnote{\url{http://www.minitlabs.com/} -- last accessed 30/11/2016}. 
%\iw{Luciano to add details on filtering}
Given this log, we discovered a business process model using the Structured BPMN Miner~\cite{AugustoCDRB16}, which showed a high level of conformance (> 99\%). After filtering out non conforming traces, we ended up with dataset that contains 5,316 traces, out of which 49 traces are distinct. These traces are based on 21 distinct event types, including one for instance creation, and have a weighted average length of 11.6 events.
%\iw{Luciano, Alex: are all event types in log6000 true events, including ``Start'' ``Process start'' and ``Process end'' ? Also, can either of you count the tasks and gateways, and put the count in the table above, please?}
%Give basic stats about this log: number of traces, number of distinct traces, number of distinct event types, average length of traces, number of attributes per event (on average), 

\subsection{Methodology and Setup}

We translated the process models into Solidity code, using the previous version of the translator from~\cite{WeberBPM2016} -- referred to as \emph{default} -- and the newly implemented translator proposed in this paper -- referred to as \emph{optimized}.
Then we compiled the Solidity code for these smart contracts into EVM bytecode and deployed them on a private Ethereum blockchain.

To assess gas cost and correctness on conformance checking, we replayed the distinct log traces against both versions of contract and recorded the results. 
%Then mention other the two other datasets we use, which come from your BPM paper. We should also perhaps give basic statistics about these logs (maybe a table can be used to give the descriptive stats of the three logs in a single go).
We hereby relied on (slightly modified versions of) the log replayer and trigger components from~\cite{WeberBPM2016}.
The replayer iterates through a log and sends the events, one by one, via a RESTful Web service call to the trigger.
The trigger accepts the service call, packages the content into a blockchain transaction and submits it. Once it observes a block that includes the transaction, it replies to the replayer with meta-data that includes block number, consumed gas, transaction outcome (accepted or failed, i.e., non-conforming), and whether the transaction completed this process instance successfully.
The modifications of these two components cater for concurrency and additional requirements from the Minit logs.

All experiments were run using a desktop PC with an Intel i5-4570 quadcore CPU without hyperthreading. Ethereum mining for our private blockchain was set to use one core.
The log replayer and the trigger ran on the same machine, interacting via the network interface with one another.
For comparability with the results reported in~\cite{WeberBPM2016}, we used the same software in the same versions that was used in those experiments, and a similar state of the blockchain as when they were run in February--March 2016.
For Ethereum mining we used the open-source software geth\footnote{\url{https://github.com/ethereum/go-ethereum/wiki/geth} -- last accessed 30/11/2016}, version v1.5.4-stable.

\subsection{Gas Costs and Correctness of Conformance Checking}

Given that gas costs and correctness of conformance checking are both deterministic, we performed a single experiment using only \emph{distinct traces}.
%Gas consumption and correctness of a trace are both deterministic: given the state of a process instance (manifested in the values of the instance contract's variables), a method call with fixed parameters, and a fixed version of the Ethereum miner software, the bytecode we generate from both translators executes in the same manner and consumes the same amount of gas. 
%Therefore, we replayed only \emph{distinct traces} for this experiment. 
For each distinct trace, we recorded the gas required for deploying an instance contract, the sum of the gas required to perform all the required contract function invocations, the number of rejected transactions due to non-conformance and the successful completion of the process instance.
%Per distinct trace we recorded the gas for deploying an instance contract, the sum of gas for all method invocations, and non-conformance faults as the number of rejected transactions and whether the process completed successfully.

\begin{wraptable}{r}{.5\textwidth}
\vspace{-9mm}
\center\tiny
\newcommand{\mywidth}{4.2em}
\begin{tabular}{|p{\mywidth}|r|l|r|r|r|}
\hline
%Dataset & \parbox[t][20pt][t]{1.15cm}{Distinct Traces} & Translator & \parbox[t][20pt][t]{1.15cm}{Correct-ness} & \parbox[t][20pt][t]{1.5cm}{Weighted avg. cost} & \parbox[t][20pt][t]{1.8cm}{\# of cheaper traces} \\
{\bf Process} & \multirow{2}{7.5mm}{\bf Tested Traces} &  \multirow{2}{0.95cm}{\bf Translat. Version} %& \multirow{2}{1.15cm}{\bf Correct-ness} 
& \multicolumn{2}{c|}{\bf W. Avg. Cost} %& \multirow{2}{1.9cm}{\bf \# of cheaper traces} 
& \multirow{2}{8.5mm}{\bf Savings (\%)}  \\
%\cline{5-6}
&&&{\bf Instant.} &{\bf Exec.} & \\
\hline
\multirow{2}{\mywidth}{Invoicing} & \multirow{2}{*}{5316*} & Default %& 100\% 
&  1,089,000 & 33,619 %& 0 
& -- \\
\cline{3-6}
&& Optimized %& 100\% 
&  807,123 & 26,093 %& 49 
& -24.97\\
\hline
\multirow{2}{\mywidth}{Supply chain} & \multirow{2}{*}{62} & Default 
%& 100\% 
& 304,084 & 25,564 %& 0 
& --\\
\cline{3-6}
&& Optimized %& 100\% 
& 298,564 & 24,744 %& 62 
& -2.48\\
\hline
\multirow{2}{\mywidth}{Incident mgmt.} & \multirow{2}{*}{124} & Default 
%& 100\% 
&  365,207 & 26,961 %& 0 
& --\\
\cline{3-6}
&& Optimized %& 100\% 
&  345,743 & 24,153 %& 124 
& -7.04\\
\hline
\multirow{2}{\mywidth}{Insurance claim} & \multirow{2}{*}{279} & Default 
%& 100\% 
& 439,143 & 27,310 %& 0 
& --\\
\cline{3-6}
&& Optimized 
%& 100\% 
& 391,510 & 25,453 %& 279 
& -8.59\\
\hline
\end{tabular}
%\medskip
\vspace{-2mm}
\caption{Gas cost experiment results}
\label{tbl:res-cost-conf}
\vspace*{-7mm}
\end{wraptable}

The results of this experiment are shown in Table~\ref{tbl:res-cost-conf}.
The base requirement was to maintain 100\% conformance checking correctness with the new translator, which we achieved. 
Our hypothesis was that the optimized translator leads to strictly monotonic improvements in cost on the process instance level. 
We tested this hypothesis by pairwise comparison of the gas consumption per trace, and confirmed it: all traces for all models incurred less cost in \emph{optimized}.
In addition to these statistics, we report the absolute costs as weighted averages, taking into account the occurrence frequencies of the distinct traces. For \emph{Invoicing,} we report the weighted average costs across the 5,316 traces; this data is obtained from a single replay of each distinct trace, multiplied by the trace occurrence frequency in the full log. % -- hence the ``*'' in the table.

%Then explain how we proceeded with the relative comparison. Explain that this experiment is done on distinct traces and why it makes sense to use distinct traces for this experiment. Maybe show old vs. new version per distinct trace, as well as weighted average taking into account number of trace occurrences.

%Give the results of the gas consumption comparison experiment.

\subsection{Throughput Experiment}

To comparatively test scalability of the approach, we analyze the throughput using the default and optimized contracts.
To this end, we used the largest of the four datasets, \emph{invoicing}, where we ordered all the events in this log chronologically, applied a cut-off at 500 complete traces, and replayed these at a high frequency.
In particular, after a ramp-up phase we ran up to all 500 process instances in parallel against the baseline, \emph{default}, and in a separate campaign against our \emph{optimized} version.
%Events within a single trace were blocking: only when the previous event from a given trace had been processed would we submit the next transaction.
%Pending transactions were submitted to the miner's transaction pool, where they waited for processing.
%%\iw{Me and Alex to discuss ramp-up, so that we don't have 500 contract creation transactions pending at the same time.}
%It should be noted, however, that the trigger is implemented in JavaScript for nodejs, and uses the RPC interface to the Geth client.
%JavaScript has limited concurrency, and hence we cannot rule out the trigger as a potential bottleneck. 
The events from the event traces are sequentially processed. The transaction for an event in a given event trace is submitted only when the transaction of its previous event gets completed. Ethereum's miner keeps a transaction pool, where pending transactions wait for being processed. It has to be noted that the trigger component is implemented in Javascript running on NodeJS. Given the limitations of the Javascript concurrency model, we cannot rule out the trigger as a potential bottleneck.
%\iw{Not sure this is the ideal place for this discussion.}

One major limiting factor for throughput is the gas limit per block: the sum of consumed gas by all transactions in a block cannot exceed this limit, which is set through a voting mechanism by the miners in the network.
To be consisted with the rest of the experimental setup, we used the block gas limit from March 2016 at approx. 4.7M gas, although the miner in its default setting has the option to increase that limit slowly by small increments. % cut this last bit?
Given the numbers in Table~\ref{tbl:res-cost-conf}, it becomes clear that this is fairly limiting:
for \emph{optimized}, instance contract creation for the invoicing dataset costs approx. 807K gas, and thus no more than 5 instances can be created within a single block; for \emph{default}, this number drops to 4.
Regular message calls cost on average 26.1K / 33.6K gas, respectively for \emph{optimized / default}, and thus a single block can contain around 180 / 140 such transactions at most. 
%The cost for \emph{default} contracts is higher, hence we hypothesized that the throughput would be lower.
These numbers do decrease further when we are not the only user of the network.

Block limit is a major consideration. However, block frequency can vary:
on the public Ethereum blockchain, mining difficulty is controlled by a formula that aims at a median inter-block time of 13-14s.
 %According to monitors such as \url{http://etherscan.io}, average inter-block time is typically between 13 and 15s.
As we have demonstrated in~\cite{WeberBPM2016}, for a private blockchain we can increase block frequency to less than a second.
Therefore, when reporting results below \emph{we use blocks as a unit of relative time}.

\begin{figure}[t!]
\center
\includegraphics[width=0.48\columnwidth]{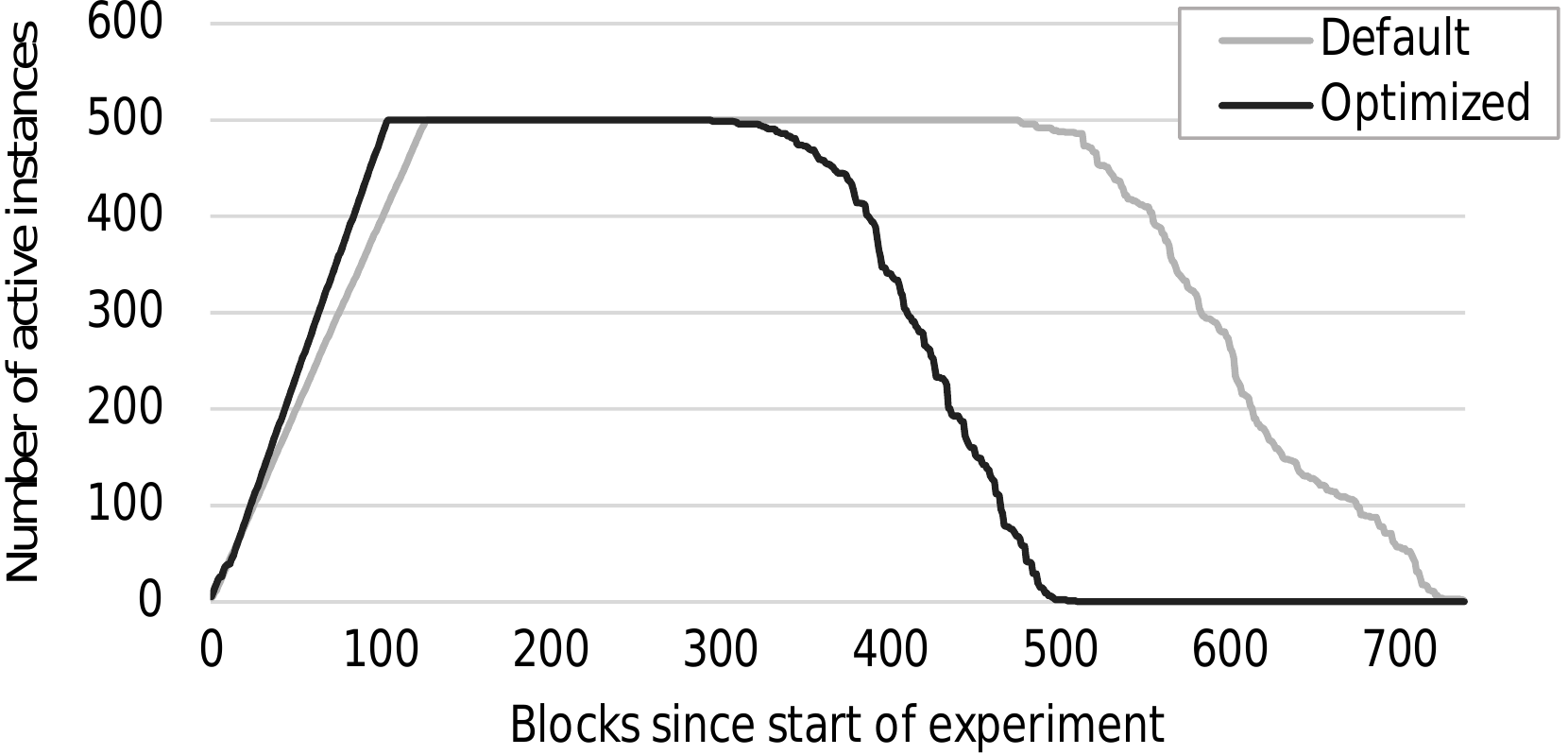}%
~~~
\includegraphics[width=0.48\columnwidth]{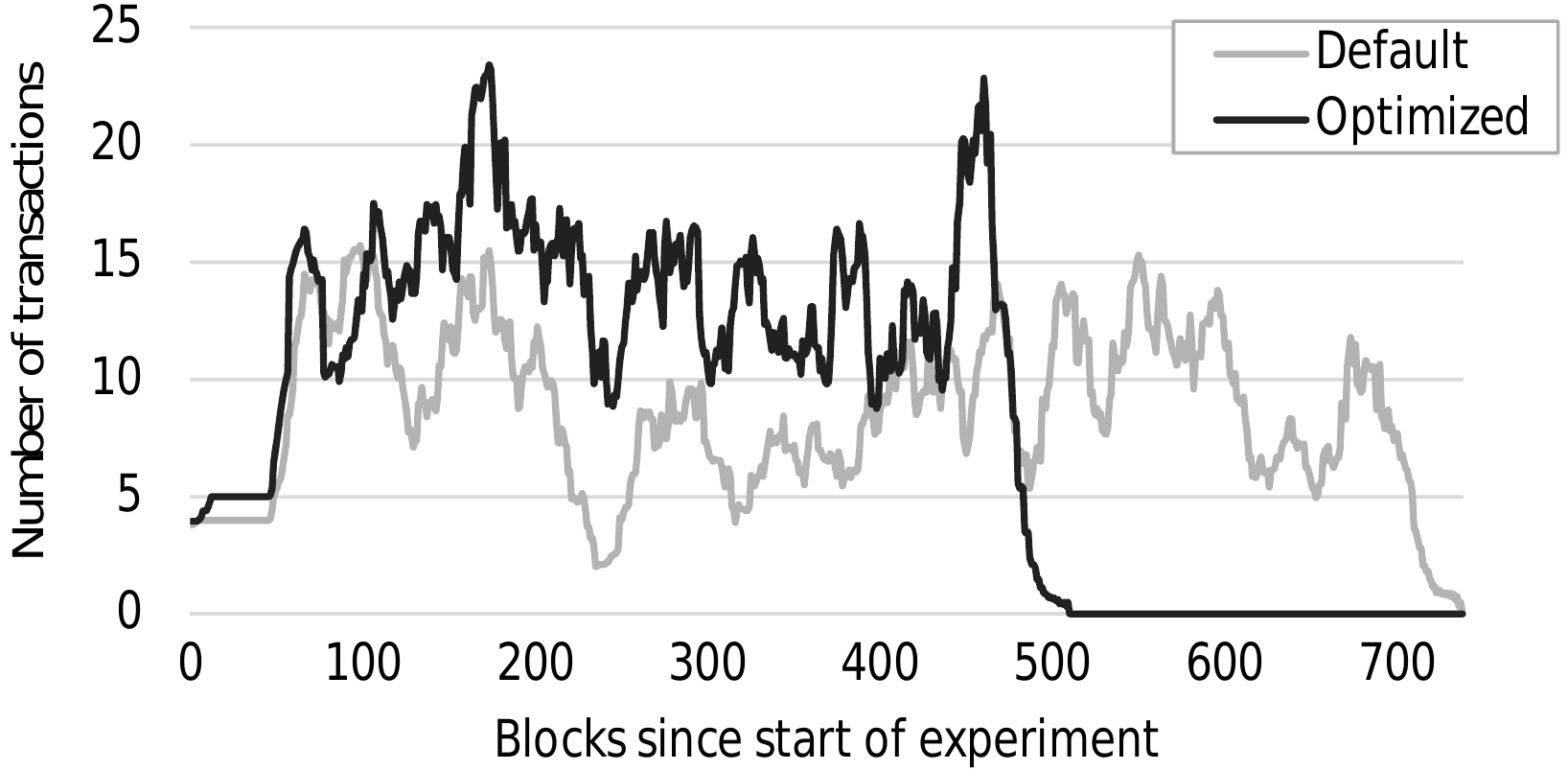}%
\vspace{-2mm}
\caption{Throughput results. Left: \# of active instances. Right: \# of transactions per block, smoothed over a 20-block time window.\label{fig:thruput}}
\vspace{-6mm}
\end{figure}

Fig.~\ref{fig:thruput} shows the main results, in terms of process instance backlog and transactions per block. Note that each datapoint in the right figure is averaged over 20 blocks for smoothing.
The main observation is that \emph{optimized} completed all 500 instances after 511 blocks, whereas \emph{default} needed 739 blocks.
The initial ramp-up phase can be seen on the right, where we see the hypothesized throughputs of 5, resp. 4, transactions per block due to the block gas limit.
As can be seen, most of the time the throughput of \emph{optimized} was higher than for \emph{default}.

\section{Conclusion}
\label{sec:conclusion}

This paper presented a method to compile a BPMN process model into a Solidity smart contract, which can be deployed on the Ethereum platform and used to enforce the correct execution of process instances. The method minimizes gas consumption by encoding the current state of the process model as a space-optimized data structure (i.e.\ a bit array with a minimized number of bits) and reducing the number of operations required to execute a process step. The experimental evaluation showed that the method significantly reduces gas consumption and achieves higher throughput relative to a previous baseline.
% by relying on the integrity mechanisms of this blockchain platform
% (i.e.\ a task completion or event occurrence)
%involving two synthetic and one real-life log shows

The presented method is a building block towards a blockchain-based collaborative business process execution engine. However, it has several limitations, including: (i) it focuses on encoding control-flow relations and data condition evaluation, leaving aside issues such as how parties in a collaboration are bound to a process instance and access control issues; (ii) it focuses on a ``core subset'' of the BPMN notation, excluding timer events, subprocesses and boundary events for example. Addressing these limitations is a direction for future work.

\bibliographystyle{splncs}
\bibliography{bibliography}

\end{document}